\documentclass[aps,prl,twocolumn,superscriptaddress,showpacs,floatfix]{revtex4-1}
\usepackage{graphicx}
\usepackage{longtable}
\usepackage{array}
\usepackage{braket}
\newcolumntype{L}{>{\centering\arraybackslash}m{2cm}}
\newcolumntype{R}{>{\centering\arraybackslash}m{1.5cm}}
\newcolumntype{K}{>{\centering\arraybackslash}m{1.3cm}}
\usepackage{gensymb}

\begin{document}

\title{Magnetic dilution and domain selection in the XY pyrochlore antiferromagnet Er$_{2}$Ti$_2$O$_7$}

\author{J.~Gaudet}
\affiliation{Department of Physics and Astronomy, McMaster University, Hamilton, ON L8S 4M1, Canada}

\author{A.~M.~Hallas}
\affiliation{Department of Physics and Astronomy, McMaster University, Hamilton, ON L8S 4M1, Canada}

\author{D. ~D.~Maharaj}
\affiliation{Department of Physics and Astronomy, McMaster University, Hamilton, ON L8S 4M1, Canada}

\author{C.~R.~C.~Buhariwalla}
\affiliation{Department of Physics and Astronomy, McMaster University, Hamilton, ON L8S 4M1, Canada}

 \author{E.~Kermarrec}
\affiliation{Department of Physics and Astronomy, McMaster University, Hamilton, ON L8S 4M1, Canada}

\author{N.~P.~Butch}
\affiliation{NIST Centre for Neutron Research, Gaithersburg Maryland USA}

\author{T.~J.~S.~Munsie}
\affiliation{Department of Physics and Astronomy, McMaster University, Hamilton, ON L8S 4M1, Canada}

\author{H.~A.~Dabkowska}
\affiliation{Brockhouse Institute for Materials Research, Hamilton, ON L8S 4M1, Canada}

\author{G.~M.~Luke}
\affiliation{Department of Physics and Astronomy, McMaster University, Hamilton, ON L8S 4M1, Canada}
\affiliation{Canadian Institute for Advanced Research, 180 Dundas Street West, Toronto, Ontario M5G 1Z8, Canada}

\author{B.~D.~Gaulin}
\affiliation{Department of Physics and Astronomy, McMaster University, Hamilton, ON L8S 4M1, Canada}
\affiliation{Brockhouse Institute for Materials Research, Hamilton, ON L8S 4M1, Canada}
\affiliation{Canadian Institute for Advanced Research, 180 Dundas Street West, Toronto, Ontario M5G 1Z8, Canada}

\date{\today}

\begin{abstract}
Below $T_N = 1.1$~K, the XY pyrochlore Er$_2$Ti$_2$O$_7$ orders into a $k=0$ non-collinear, antiferromagnetic structure referred to as the $\psi_2$ state. The magnetic order in Er$_2$Ti$_2$O$_7$ is known to obey conventional three dimensional (3D) percolation in the presence of magnetic dilution, and in that sense is robust to disorder. Recently, however, two theoretical studies have predicted that the $\psi_2$ structure should be unstable to the formation of a related $\psi_3$ magnetic structure in the presence of magnetic vacancies. To investigate these theories, we have carried out systematic elastic and inelastic neutron scattering studies of three single crystals of Er$_{2-x}$Y$_x$Ti$_2$O$_7$ with $x=0$ (pure), 0.2 (10$\%$-Y) and 0.4 (20$\%$-Y), where magnetic Er$^{3+}$ is substituted by non-magnetic Y$^{3+}$. We find that the $\psi_2$ ground state of pure Er$_2$Ti$_2$O$_7$ is significantly affected by magnetic dilution. The characteristic domain selection associated with the $\psi_2$ state, and the corresponding energy gap separating $\psi_2$ from $\psi_3$, vanish for Y$^{3+}$ substitutions between 10$\%$-Y and 20$\%$-Y, far removed from the 3D percolation threshold of $\sim$60$\%$-Y. The resulting ground state for Er$_2$Ti$_2$O$_7$ with magnetic dilutions from 20$\%$-Y up to the percolation threshold is naturally interpreted as a frozen mosaic of $\psi_2$ and $\psi_3$ domains.
\end{abstract}


\maketitle

The network of corner-sharing tetrahedra that make up the pyrochlore lattice is one of the canonical architectures for geometrical frustration in three dimensions~\cite{RevModPhys.82.53}. In the pyrochlore magnets with chemical composition $A_2B_2$O$_7$, each of the $A^{3+}$ and $B^{4+}$ sublattices independently form such a network, that can be decorated by many ions. The rare earth titanate family, $R_2$Ti$_2$O$_7$, which features a single magnetic $R^{3+}$ site and non-magnetic Ti$^{4+}$ on the $B$-site, has been of specific interest. This family displays a great variety of exotic magnetic ground states, including classical~\cite{harris1997geometrical,ramirez1999zero,castelnovo2008magnetic} and quantum spin ice~\cite{Ross2011,gingras2014quantum,applegate2012vindication,lee2012generic}, as well as spin liquid states~\cite{gardner1999cooperative,takatsu2011quantum,gingras2000thermodynamic}. Such exotic states arise from different combinations of magnetic anisotropies with differing exchange and dipolar interactions, depending on the nature of the $R^{3+}$ magnetic ion.

 \begin{figure*}[tbp]
\linespread{1}
\par
\includegraphics[width=7in]{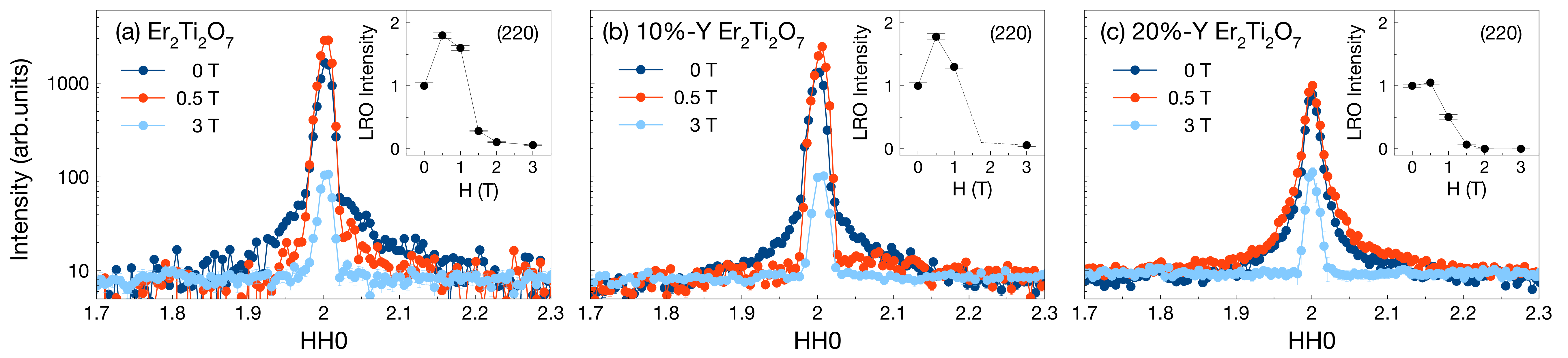}
\par
\caption{Magnetic Bragg peak intensity for $Q$ = (220) as a function of field for the pure, 10$\%$-Y and 20$\%$-Y diluted samples of Er$_{2-x}$Y$_x$Ti$_2$O$_7$. The elastic scattering was isolated by integrating from $-0.2$ to 0.2 in the [00L] direction and from $-0.1$ to 0.1 meV in energy transfer. The insets show the integrated intensity of the long range ordered component of the scattering for each of the three samples. Examples of the fits used to obtain these intensities are shown in Fig.1 of the Supplemental Material. The dashed line in the inset of (b) indicates the expected behavior for the 10$\%$-Y sample at 1.75~T.}
\label{ElasCut} 
\end{figure*}

Much current interest is focused on the XY antiferromagnetic pyrochlore Er$_2$Ti$_2$O$_7$, which displays an antiferromagnetic Curie-Weiss susceptibility with $\theta_{CW}= -22$~K~\cite{PhysRevB.68.020401}. Crystal field effects on Er$^{3+}$ in this environment give rise to a $g$-tensor with XY anisotropy~\cite{Cao2009,Bonville2013,bertin2012crystal}. In contrast to the exotic ground states displayed by other members of the $R_2$Ti$_2$O$_7$ family, Er$_2$Ti$_2$O$_7$ has a rather conventional $k=0$, $\Gamma_5$ antiferromagnetic ground state below $T_N=1.1$~K~\cite{PhysRevB.68.020401,Poole2007,Lago2005,Dalmas2012,blote1969heat}. $\Gamma_5$ is made up of two basis vectors, referred to as $\psi_2$ and $\psi_3$, and sophisticated magnetic crystallography was used to identify the ordered structure in Er$_2$Ti$_2$O$_7$ as $\psi_2$~\cite{Poole2007}. The microscopic spin Hamiltonian for Er$_2$Ti$_2$O$_7$ was determined through measurements of the spin wave dispersions within its field polarized state \cite{PhysRevLett.109.167201,PhysRevB.90.060410}. This work, amongst others \cite{PhysRevLett.109.077204,Wong2013,PhysRevB.68.020401}, has provided an explanation for the selection of $\psi_2$ within the $\Gamma_5$ manifold in terms of a quantum order-by-disorder mechanism. 
The order-by-disorder scenario has been further strengthened by works showing that thermal fluctuations also select $\psi_2$ over $\psi_3$~\cite{PhysRevB.88.220404,Zhitomirsky2014,Javanparast2015,McClarty2014}. A spin wave gap of $0.053 \pm 0.006$~meV has been successfully measured in Er$_2$Ti$_2$O$_7$, consistent with the order-by-disorder mechanism~\cite{PhysRevLett.112.057201}. Recently, there has been a proposal that the selection of the $\psi_2$ ordered state could also occur via higher multipolar interactions, originating from virtual crystal field transitions~\cite{Rau2015,PhysRevB.90.060410}.\

The phase transition to the $\psi_2$ antiferromagnetic ordered state in Er$_2$Ti$_2$O$_7$ is unusual amongst the $R_2$Ti$_2$O$_7$ family by virtue of how conventional it is. In contrast to Yb$_2$Ti$_2$O$_7$~\cite{PhysRevB.86.174424,Gaudet2015,PhysRevB.84.172408} and Tb$_2$Ti$_2$O$_7$~\cite{PhysRevB.87.060408,PhysRevB.92.245114}, there appears to be no sample dependence to its low temperature phase diagram. Also, systematic studies of magnetic dilution at the rare earth site in Er$_2$Ti$_2$O$_7$ are consistent with three dimensional (3D) percolation theory~\cite{Niven2014}. It was therefore surprising that two recent, independent theoretical works predicted that, upon magnetic dilution, the $\psi_2$ ground state selection in Er$_2$Ti$_2$O$_7$ should be unstable to the selection of its $\psi_3$ partner within the $\Gamma_5$ manifold~\cite{PhysRevB.90.094412,PhysRevB.91.064401}.\

Such a possible $\psi_2$ to $\psi_3$ phase transition upon magnetic dilution in Er$_2$Ti$_2$O$_7$ has yet to be explored experimentally and is the topic of this manuscript. The phase transition between $\psi_2$ and $\psi_3$ in Er$_2$Ti$_2$O$_7$ is predicted to occur near a critical dilution of $\sim$10\%~\cite{PhysRevB.90.094412,PhysRevB.91.064401}. As Y$^{3+}$ is non-magnetic and comparable in size to Er$^{3+}$, it is an ideal ion to employ for magnetic dilution studies. Hence, we have grown single crystal samples of Er$_{2-x}$Y$_{x}$Ti$_2$O$_7$ with $x=0$ (pure), 0.2 (10$\%$-Y) and 0.4 (20$\%$-Y) to investigate a possible change in the magnetic ground state. We performed time-of-flight inelastic neutron scattering measurements on these three single crystals at various temperatures and magnetic fields. Our results show that the signature for $\psi_2$ long range magnetic order disappears upon dilution. Instead, we observe a softening of the zero field, low energy excitations at the (220) ordering wavevector, consistent with a closing of the spin gap. At a relatively high level of magnetic dilution, on the order of 20$\%$-Y, the system cannot globally select the $\psi_2$ state over $\psi_3$, and we suggest that it forms a frozen mosaic of both $\psi_2$ and $\psi_3$ domains.\

\begin{table}[tbp]
\begin{tabular}{| l || >{\centering}p{1.2cm} | p{1.75cm}<{\centering} | p{2cm}<{\centering}|}
\toprule
 & $T_N$(K) & $\theta_{\text{CW}}$(K) & $\mu_{\text{eff}}$/Er$^{3+}$ \\
\colrule
Er$_2$Ti$_2$O$_7$(0\%) & 1.1(1)  & $-18.0(2)$~ & 9.555(4)~$\mu_{\text{B}}$ \\
Er$_{1.8}$Y$_{0.2}$Ti$_2$O$_7$(10\%) & 1.04(5) & $-17.2(3)$~ & 8.965(7)~$\mu_{\text{B}}$  \\
Er$_{1.6}$Y$_{0.4}$Ti$_2$O$_7$(20\%) & 0.74(8) & $-17.7(3)$~ & 8.523(6)~$\mu_{\text{B}}$ \\
\botrule
\end{tabular}
\caption{Summary of the dc susceptibility measurements performed on each of Er$_{2-x}$Y$_x$Ti$_2$O$_7$ with 0\%, 10\% and 20\% yttrium-doped samples. The full data are shown in the Supplemental Material.
} 
\label{CurieWeiss}
\end{table} 

The three Er$_{2-x}$Y$_{x}$Ti$_2$O$_7$ samples were characterized by dc magnetic susceptibility, and these results are summarized in Table~\ref{CurieWeiss}. In each case, the transition to the ordered phase is marked by a cusp and a bifurcation of the field cooled and zero field cooled susceptibilities, as shown in the Supplemental Material~\cite{SuppEYTO}. Both the N\'eel temperature and the paramagnetic moment per formula unit systematically decrease as a function of the magnetic dilution. The Curie-Weiss temperature ($\theta_{\text{CW}}$) does not vary appreciably in these three materials, ranging from $-17.2(3)$~K to $-18.0(2)$~K.\

We performed time-of-flight elastic and inelastic neutron scattering on these three single crystals using the Disc Chopper Spectrometer at the NIST Center for Neutron Research~\cite{DCS}. Measurements were performed in a dilution refrigerator in both zero magnetic field and an applied magnetic field along the [1,-1,0] direction, perpendicular to the (HHL) scattering plane. For each of the three samples, magnetic Bragg peaks form below their respective N\'eel transitions (see Table~\ref{CurieWeiss}). The relative intensities of the magnetic Bragg reflections in the 10$\%$-Y and 20$\%$-Y crystals are unchanged from pure Er$_2$Ti$_2$O$_7$. That is to say, all three order into the $\Gamma_5$ irreducible representation, which is dominated by an intense magnetic Bragg reflection at (220). In large magnetic fields, all three enter a field polarized state, characterized by an intense (111) magnetic Bragg reflection and the disappearance of the (220) magnetic reflection~\cite{Ruff2008,McClarty2009}.\

The $\Gamma_5$ irreducible representation is comprised of two basis vectors, $\psi_2$ and $\psi_3$, which are indistinguishable in an unpolarized elastic neutron scattering experiment. However, the distinction between $\psi_2$ and $\psi_3$ ground state selection can be resolved by considering the magnetic field dependence of the (220) magnetic Bragg peak. Both $\psi_2$ and $\psi_3$ display six distinct domains that are degenerate in zero field. This degeneracy is lifted by the application of a small magnetic field along the [1,-1,0] direction~\cite{PhysRevLett.109.167201,Dun2015,maryasin2015low}. For $\psi_2$, the two domains selected in a magnetic field are the ones that maximize the (220) magnetic Bragg peak, resulting in a twofold increase of its intensity. Domain selection in a [1-10] field for the $\psi_3$ state is less clear. Two scenarios are proposed; one selecting two domains which reduce the intensity of (220)~\cite{PhysRevLett.109.167201,Dun2015} while the other selects four domains that resulting in a factor of 1.5 increase for (220)~\cite{maryasin2015low,Zhitomirskyprivate}. We note these domain effects occur in small [1,-1,0] fields, significantly below the onset of the field polarized state. The effect of the canting angle on the (220) Bragg peak intensity have been well characterized for Er$_2$Ti$_2$O$_7$~\cite{Ruff2008,McClarty2009,Cao2010} and little effect is seen on it's intensity between 0.1T and 1T. Hence, we used a field of 0.5T to investigate domain effects, which, is sufficiently large to overcome pinning of domains by disorder in a small field.

\begin{figure*}[tbp]
\linespread{1}{}
\par
\includegraphics[width=7in,height=2.3in]{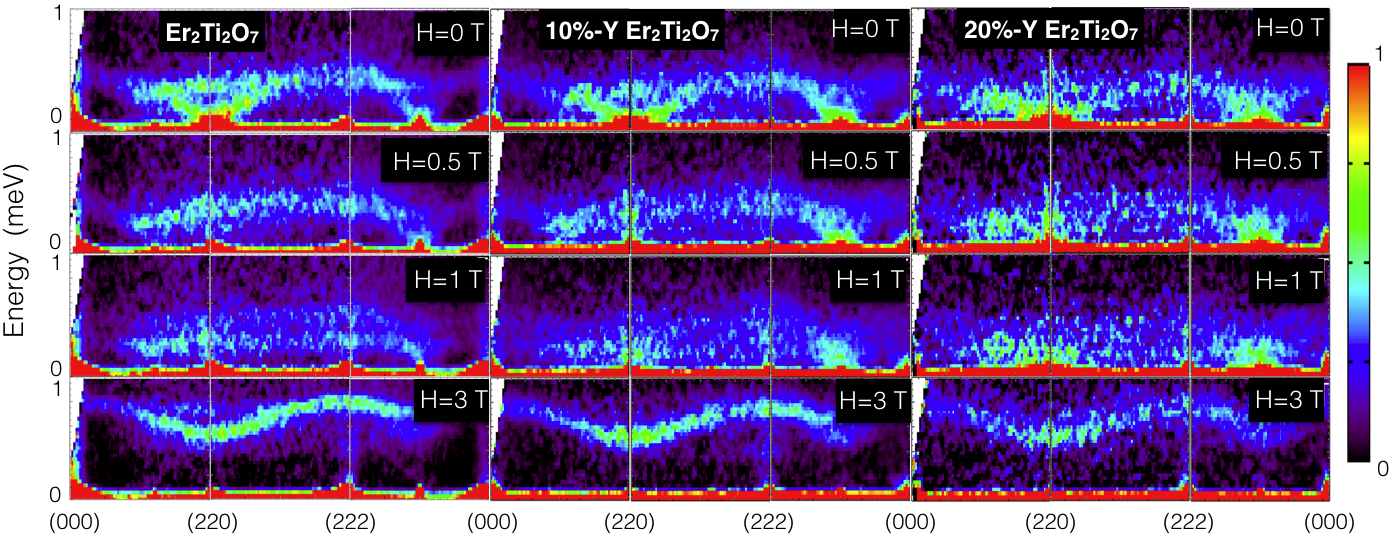}
\par
\caption{Spin wave spe{}ctra along three different directions in the HHL plane: [HH0], [22L], and [HHH], all at $T=$0.1~K. The spectra are shown for each of the three samples in zero magnetic field, and with magnetic fields of 0.5~T, 1~T, and 3~T applied along the [1-10] direction. The contour plots along the [HH0], [22L] and [HHH] directions are obtained by integrating the 3D data set, $S(\vec{Q},E)$, over $ -0.2 \leq [00L] \leq 0.2 $, $1.8 \leq [HH0] \leq 2.2$ and $0.2 \leq [HH-2H] \leq 0.2$, respectively.}
\label{SpinWaveField}
\end{figure*}

We first consider the elastic scattering from the three crystals at the (220) Bragg position in zero field, 0.5~T and 3~T, as shown in Fig.~\ref{ElasCut}. The intensities have been normalized using the Bragg intensity at H~=~3T, which corresponds to the polarized state where the scattered intensity at (220) is purely structural. As the field is increased to 0.5~T, we observe a marked increase in the peak intensity for the pure sample as well as for the 10$\%$-Y doped sample, but not for the 20$\%$-Y diluted crystal. More striking is the evolution of the shape of the (220) Bragg peak with magnetic dilution. In zero field, the undoped sample exhibits a resolution-limited Bragg peak with a small amount of diffuse scattering extending out along [HH0] (Fig.~\ref{ElasCut}(a)). Upon dilution, the relative contribution of the resolution-limited Bragg scattering weakens with a corresponding increase in the diffuse scattering, such that the lineshape of the 20$\%$-Y sample is dominated by a Lorentzian form in both zero field and at 0.5~T (Fig.~\ref{ElasCut}(c)). 

To determine the origin of the diffuse scattering around (220), we look to the full inelastic neutron spectra in Fig.~\ref{SpinWaveField}. These data are shown along three reciprocal space directions for each of the three samples in four different magnetic fields applied along the [1,-1,0] direction. In zero field (top row of Fig.~\ref{SpinWaveField}), the spin waves of the three samples are qualitatively similar, consisting of one flat branch near 0.4~meV and quasi-Goldstone modes which soften at (111) and (220). A previous high resolution inelastic neutron scattering study of the pure sample determined that the quasi-Goldstone modes are gapped by 0.053 $\pm$ 0.006 meV~\cite{PhysRevLett.112.057201}. As the energy resolution associated with the present inelastic measurements is 0.09 meV, elastic cuts of the form shown in Fig.~\ref{ElasCut} necessarily integrate over some of the spectral weight of these quasi-Goldstone modes. Thus, for the pure sample, this low energy inelastic scattering is the origin of the diffuse scattering observed in our cuts over (220) in Fig.~\ref{ElasCut}. However for the 10$\%$-Y and especially the 20$\%$-Y sample, our data suggests that the spin gap is reduced from 0.053 meV, allowing for additional quasi-elastic and elastic magnetic scattering, characteristic of a frozen mosaic of $\psi_2$ and $\psi_3$ domains.

In order to understand the contributions to the scattering in the cuts of Fig.~\ref{ElasCut}, we fit each elastic data set to the sum of a Gaussian and a Lorentzian lineshape, quantifying the magnetic long-range order (LRO) and the dynamic, quasi-elastic or frozen spin contributions, respectively. An example of such a fit for each sample is shown in the Supplemental Material~\cite{SuppEYTO}. The resulting fits show that the relative contribution of the diffuse, Lorentzian lineshape grows as a function of doping and accounts for $\sim$75$\%$ of the $Q$-integrated scattering near (220) in the 20$\%$-Y sample at $T=0.1$~K and zero field. We therefore suggest that with increasing magnetic dilution, $x$, the spin excitations near (220) in Er$_{2-x}$Y$_x$Ti$_2$O$_7$ soften to lower energies and freeze. This is likely the result of a collapsing spin gap, a direct measure of the selection of $\psi_2$ over $\psi_3$.\

We can now isolate the LRO component of the elastic scattering and study its field dependence at $T=0.1$~K. Once again, we use the fits to the scattering around (220), wherein a resolution-limited Gaussian lineshape represents the LRO and a broadened Lorentzian represents the dynamic, quasi-elastic and frozen spin response captured by our finite energy resolution. The LRO integrated intensity at (220) is shown as a function of a [1,-1,0] magnetic field for each sample in the insets of Fig.~\ref{ElasCut}. Comparing the LRO intensity at zero field and 0.5~T, we observe a twofold increase for both the pure and 10$\%$-Y samples, while the LRO is unchanged for the 20$\%$-Y sample. This indicates, via the domain selection scenario described above, that the pure and 10$\%$-Y samples order into $\psi_2$, but the 20$\%$-Y sample does not. Furthermore, these results are inconsistent with the detailed theoretical predictions that low levels($\sim$7$\%$) of magnetic dilution should induce a transition from $\psi_2$ to $\psi_3$~\cite{PhysRevB.90.094412,PhysRevB.91.064401}.

\begin{figure}[tbp]
\linespread{1}
\par
\includegraphics[width=3.3in]{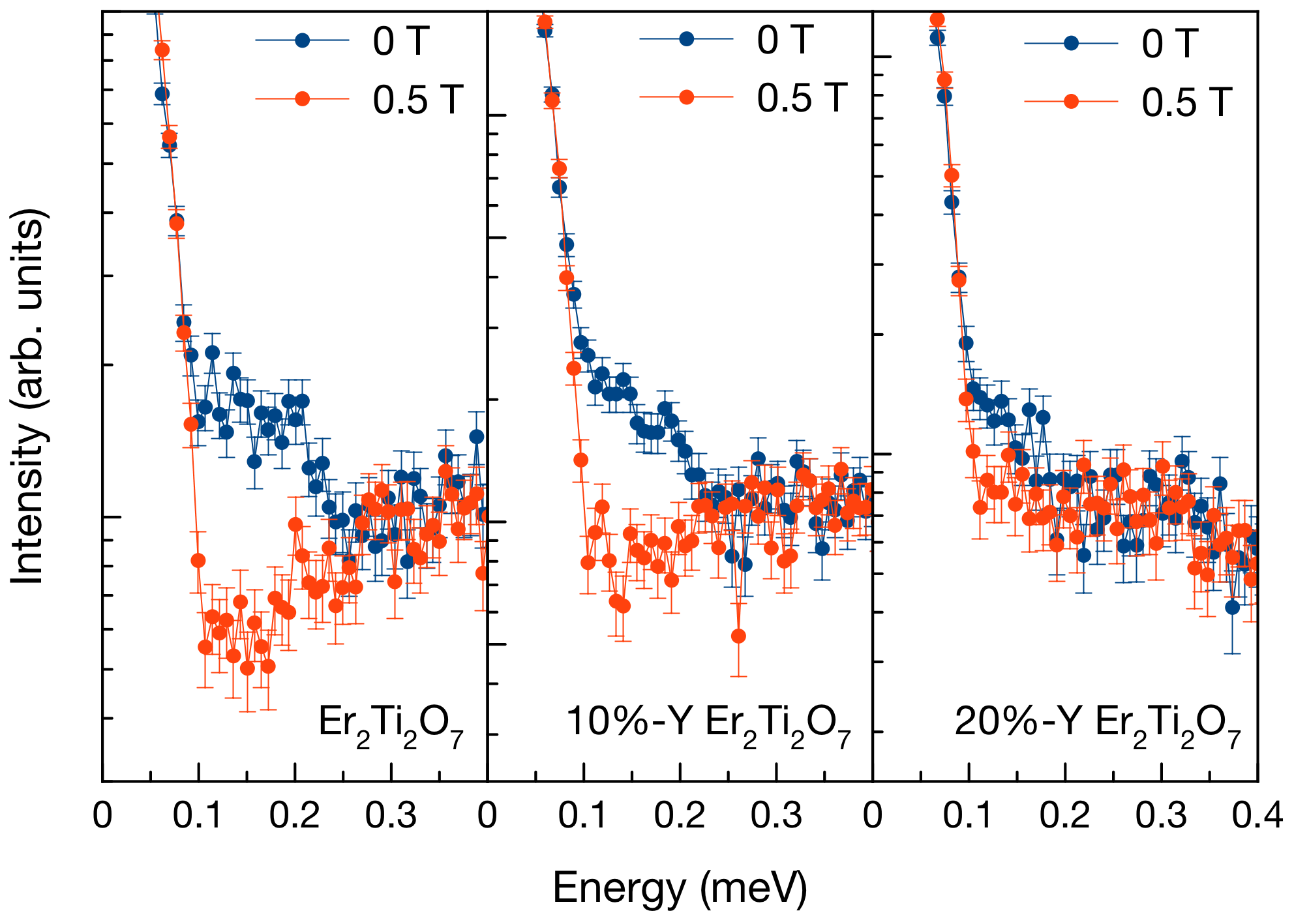}
\par
\caption{Inelastic intensity as a function of energy at the (220) Bragg position for the pure, 10$\%$-Y and 20$\%$-Y diluted samples of Er$_{2-x}$Y$_x$Ti$_2$O$_7$. These plots are obtained by integrating the 3D data sets $S(\vec{Q},E)$ at $T=0.1$~K over $1.6 \leq [HH0] \leq 2.4$ and $ -0.5 \leq [00L] \leq 0.5$.}
\label{220doping}
\end{figure}

\begin{figure}[tbp]
\linespread{1}
\par
\includegraphics[width=3.3in]{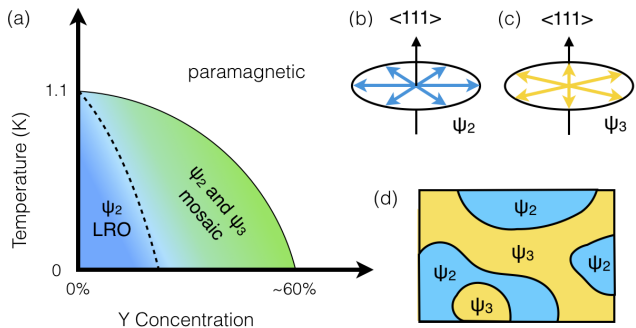}
\par
\caption{(a) The schematic temperature-dilution phase diagram for Er$_{2-x}$Y$_x$Ti$_2$O$_7$. A dilution between 10$\%$-Y and 20$\%$-Y marks a crossover from a pure $\psi_2$ state into a frozen mosaic of $\psi_2$ and $\psi_3$ domains. The six discrete domains allowed by (b) $\psi_2$ and (c) $\psi_3$ within the XY plane. (d) Schematic illustration of the state at moderate dilution showing the mosaic of $\psi_2$ and $\psi_3$ domains.}
\label{PhaseDiag}
\end{figure}

A striking feature of Fig.~\ref{ElasCut}, is the almost complete absence of diffuse scattering around (220) at $H=0.5$~T for both the pure and 10$\%$-Y samples, but not the 20$\%$-Y sample. Examination of the detailed spin wave spectra in Fig.~\ref{SpinWaveField}, shows that a 0.5~T field gaps the quasi-Goldstone mode at (220) in the pure and 10$\%$-Y samples (second row of Fig.~\ref{SpinWaveField}). Meanwhile, quasi-elastic scattering persists around (220) in the 20$\%$-Y sample to fields greater than 1~T. To quantify this effect, we integrated a small portion of reciprocal space around (220) and plotted the energy dependence of this scattering in zero field and 0.5~T for all three samples (Fig.~\ref{220doping}). It is clear that the quasi-elastic scattering from the quasi-Goldstone modes in the pure and 10$\%$-Y samples are gapped by $\sim$0.3 meV in a 0.5~T field. However, for the 20$\%$-Y sample, the zero field excitations are soften and the application of a 0.5~T magnetic field makes little difference to the low energy spectral weight near (220). Both of these observations are consistent with the absence of a spin wave gap and the associated domain selection in a magnetic field in the 20$\%$-Y sample .

We propose the schematic phase diagram shown in Fig.~\ref{PhaseDiag} to describe the magnetic state in magnetically diluted Er$_{2-x}$Y$_x$Ti$_2$O$_7$. The phase transition and $\psi_2$ ground state selection are already well-established in the pure material~\cite{Poole2007,PhysRevB.68.020401}, along with the concomitant opening of a $0.053 \pm 0.006$~meV spin wave gap at $T_N = 1.1$~K~\cite{PhysRevLett.112.057201}. Heat capacity measurements as a function of magnetic dilution are consistent with conventional 3D percolation theory~\cite{Gaulin2004}, and a percolation threshold near 60$\%$-Y has also been established~\cite{Niven2014}. The present measurements on magnetically diluted single crystal samples shows that $\psi_2$ domain selection is observed at low temperatures for the 10$\%$-Y sample but not for the 20$\%$-Y sample. This strongly suggests the presence of a phase boundary that mirrors the collapse of the spin gap as a function of dilution, indicated by the dashed line in Fig.~\ref{PhaseDiag}(a). Weak dilution to the left of the dashed line produces a stable set of $\psi_2$ domains (Fig.~\ref{PhaseDiag}(b)), at least above our minimum $T=0.1$~K. However, to the right of the dashed line, the spin gap collapses and there is no mechanism for selection of $\psi_2$ over $\psi_3$. A ground state characterized by a linear combination of $\psi_2$ and $\psi_3$ forming the full U(1) manifold is consistent with our data. However, in light of the theoretical study of ref~\cite{PhysRevB.90.094412,PhysRevB.91.064401}, it is then natural to think of the diluted system forming a frozen mosaic of $\psi_2$ and $\psi_3$ domains, with the $\psi_3$ domains (Fig.~\ref{PhaseDiag}(c)) pinned by locally high concentrations of the quenched vacancies (Fig.~\ref{PhaseDiag}(d)). It is interesting to note that a similar mixed $\psi_2$ and $\psi_3$ state is also observed in NaCaCo$_2$F$_7$ but believed to originate from bond disorder~\cite{Ross2016}.

We find that the $\psi_2$ ordered state in Er$_2$Ti$_2$O$_7$ displays remarkable fragility induced by the presence of magnetic vacancies. Measurements as a function of temperature in our most magnetically dilute sample, 20$\%$-Y, provides no evidence for an additional phase transition between $T_N = 0.74(8)$~K and our base temperature of 0.1~K, as shown in Fig. 3 of the Supplemental Material~\cite{SuppEYTO}. Nonetheless, we do identify a change in phase behavior between $\psi_2$ and a ground state characterized by no selection of $\psi_2$ over $\psi_3$, likely a frozen mosaic of the two, all at dilution concentrations far below the 3D percolation threshold. Clearly, the preceding theoretical work correctly identified the sensitivity of the $\psi_2$ ground state selection to the presence of this form of quenched disorder, if not the detailed manifestation of the disorder on the phase behavior. As such, we hope our characterization of the spin statics and dynamics in single crystal Er$_{2-x}$Y$_x$Ti$_2$O$_7$, and low temperature phase behavior as a function of dilution, will motivate a complete understanding of these exotic, and fragile, ordered states. 

\begin{acknowledgments}
We wish to acknowledge useful conversations with Leon Balents, Kate Ross, Mike Zhitomirsky, Michel Gingras, and Jeff Rau. We would also like to thank Juscelino Leao for his assistance with the sample environment. This work was supported by the Natural Sciences and Engineering Research Council of Canada and the Canada Foundation for Innovation. Work at the NIST Center for Neutron Research is supported in part by the National Science Foundation under Agreement No. DMR-0944772.
\end{acknowledgments}

\bibliography{EYTO}

\end{document}

\title{SUPPLEMENTAL MATERIAL: \\
Magnetic dilution and domain selection in the XY pyrochlore antiferromagnet Er$_{2}$Ti$_2$O$_7$}

\author{J.~Gaudet}
\affiliation{Department of Physics and Astronomy, McMaster University, Hamilton, ON L8S 4M1, Canada}

\author{A.~M.~Hallas}
\affiliation{Department of Physics and Astronomy, McMaster University, Hamilton, ON L8S 4M1, Canada}

\author{D.~D.~Maharaj}
\affiliation{Department of Physics and Astronomy, McMaster University, Hamilton, ON L8S 4M1, Canada}

\author{C.~R.~C.~Buhariwalla}
\affiliation{Department of Physics and Astronomy, McMaster University, Hamilton, ON L8S 4M1, Canada}

 \author{E.~Kermarrec}
\affiliation{Department of Physics and Astronomy, McMaster University, Hamilton, ON L8S 4M1, Canada}

\author{N.~P.~Butch}
\affiliation{NIST Centre for Neutron Research, Gaithersburg Maryland USA}

\author{T.~J.~S.~Munsie}
\affiliation{Department of Physics and Astronomy, McMaster University, Hamilton, ON L8S 4M1, Canada}

\author{H.~A.~Dabkowska}
\affiliation{Brockhouse Institute for Materials Research, Hamilton, ON L8S 4M1, Canada}

\author{G.~M.~Luke}
\affiliation{Department of Physics and Astronomy, McMaster University, Hamilton, ON L8S 4M1, Canada}
\affiliation{Canadian Institute for Advanced Research, 180 Dundas Street West, Toronto, Ontario M5G 1Z8, Canada}

\author{B.~D.~Gaulin}
\affiliation{Department of Physics and Astronomy, McMaster University, Hamilton, ON L8S 4M1, Canada}
\affiliation{Brockhouse Institute for Materials Research, Hamilton, ON L8S 4M1, Canada}
\affiliation{Canadian Institute for Advanced Research, 180 Dundas Street West, Toronto, Ontario M5G 1Z8, Canada}

\date{\today}

\maketitle

\section{Magnetic susceptibility}

\begin{figure}[tbp]
\linespread{1}
\par
\includegraphics[width=3.1in]{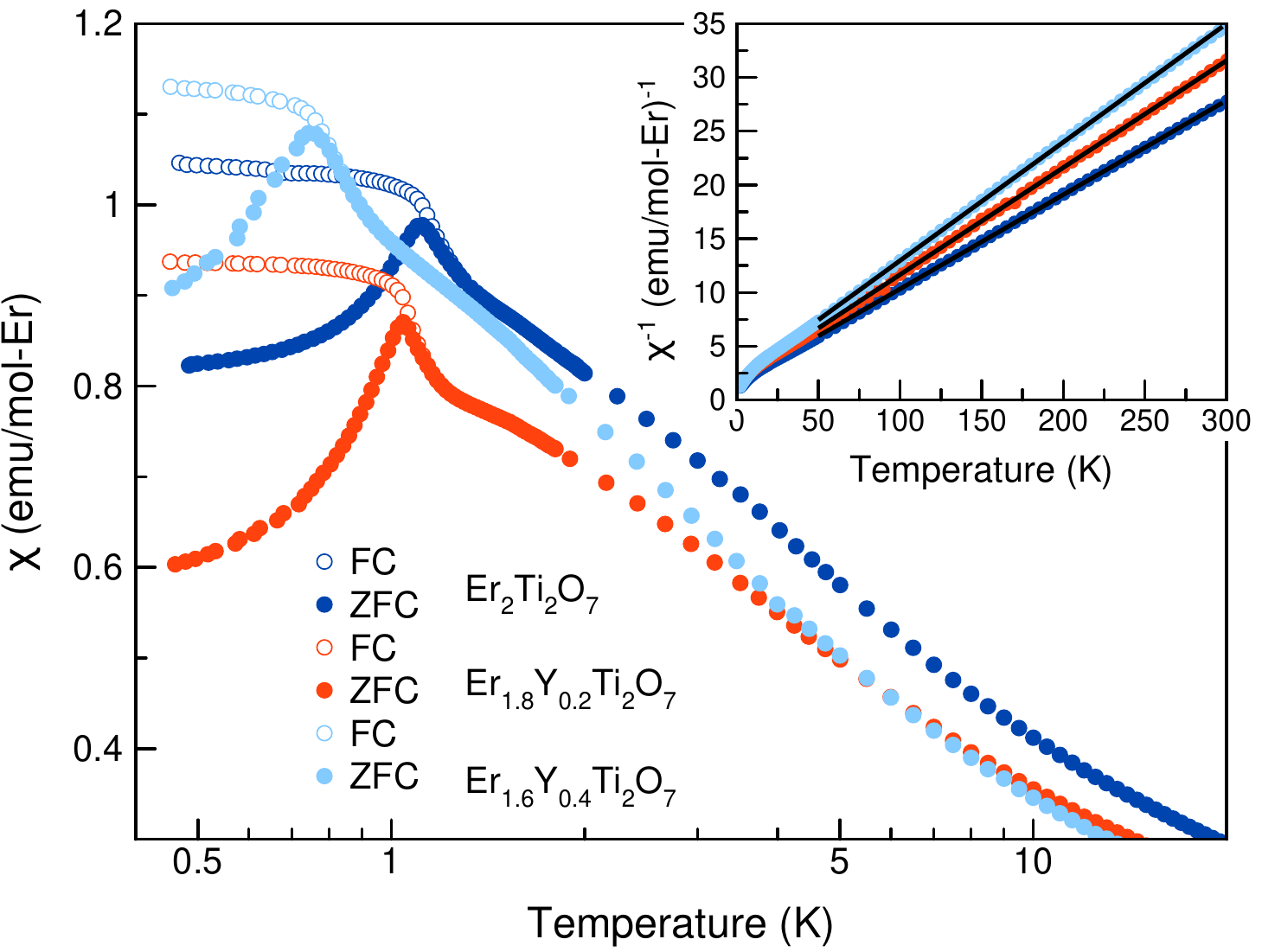}
\par
\caption{Magnetic susceptibility data obtained for Er$_{2-x}$Y$_{x}$Ti$_2$O$_7$ with $x=$0 (pure), 0.2 (10$\%$-Y) and 0.4 (20$\%$-Y). Powder samples from crushed single crystals used in the neutron scattering measurements were employed.  For each sample, both the zero field cooled (ZFC - closed symbols) and the field cooled (FC - open symbols) susceptibility are shown. The inset shows the Curie-Weiss fits performed for each sample.}
\label{MagSus}
\end{figure}

Single crystals of Er$_{2-x}$Y$_{x}$Ti$_2$O$_7$ with $x=0$ (pure), 0.2 (10$\%$-Y) and 0.4 (20$\%$-Y) were grown at McMaster University using a two-mirror floating zone image furnace. The growth procedure closely resembled the one used in Ref.~\cite{Niven2014,balakrishnan1998single,Bryan2012}. The single crystal of Er$_2$Ti$_2$O$_7$ used for this study is the same as the one studied by previous neutron scattering work~\cite{Ruff2008,PhysRevLett.109.167201,PhysRevLett.112.057201}. The susceptibility measurements were performed using a Quantum Design MPMS magnetometer equipped with a $^{3}$He insert and the results are shown in Fig.~\ref{MagSus}.

Magnetic susceptibility measurements were performed on a small portion of the crushed single crystal of undoped Er$_2$Ti$_2$O$_7$ and on crushed single crystal of the 10$\%$-Y and 20 $\%$-Y samples. The dc susceptibility was measured on warming using both a field cooled and a zero field cooled protocol with an external field of 0.01~T. In each case, the transition to long range magnetic order is marked by a cusp in the susceptibility as well as a bifurcation of the field cooled and zero field cooled susceptibilities. As the cusp and the bifurcation in the susceptibility do not occur concomitantly, we define $T_N$ as the cusp maxima, and the difference between the two as the error bar. The undoped Er$_2$Ti$_2$O$_7$ orders at $T_N = 1.1(1)$~K, consistent with previous studies~\cite{PhysRevB.68.020401}. The effect of diluting magnetic Er$^{3+}$ with non-magnetic Y$^{3+}$ is to suppress the magnetic ordering transition to lower temperature. The 10$\%$-Y and 20$\%$-Y samples have N\'eel temperatures of $T_N = 1.04(5)$~K and $T_N = 0.74(8)$~K, respectively. This suppression of $T_N$ is consistent with previous studies of yttrium-dilution in Er$_2$Ti$_2$O$_7$~\cite{Niven2014}.

At high temperatures, Curie-Weiss behavior is observed in both undoped and yttrium-doped Er$_2$Ti$_2$O$_7$. Curie-Weiss fits were performed for each sample between 50~K and 300~K, and the results of these fits are summarized in Table I of the main manuscript. The Curie-Weiss temperature ($\theta_{\text{CW}}$) does not change appreciably in these three materials, ranging from $-17.2(3)$~K to $-18.0(2)$~K. The paramagnetic moment extracted from the high-temperature Curie-Weiss fits, 9.555(4)$\mu_B$ for Er$^{3+}$ as well as prior bulk magnetization studies~\cite{Bramwell2000}. However, the paramagnetic moment per Er$^{3+}$ systematically decreases with yttrium doping, likely a consequence of differences in the high energy crystal electric field schemes in these three materials. We note, however, that our inelastic scattering study revealed no low-lying crystal electric field levels in either the 10$\%$-Y or 20$\%$-Y samples up to 3 meV.

\section{Long range ordered and diffuse contributions to the (220) Bragg peak in E\lowercase{r}$_{2-x}$Y$_x$T\lowercase{i}$_2$O$_7$}

\begin{figure*}[tbp]
\linespread{1}
\par
\includegraphics[width=7in]{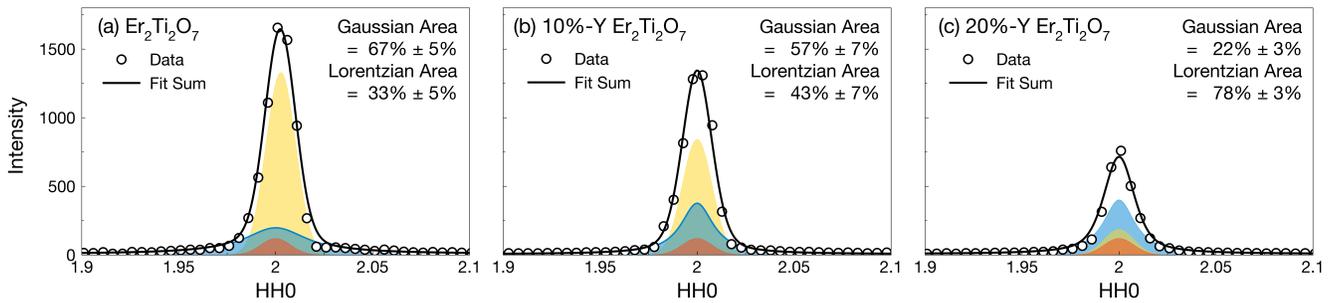}
\par
\caption{Typical fits of the (220) magnetic Bragg peak performed in this work for Er$_{2-x}$Y$_x$Ti$_2$O$_7$ with $x=0$ (pure), 0.2 (10$\%$-Y) and 0.4 (20$\%$-Y)) at $T=0.1$~K in zero field. The yellow areas represent the elastic, resolution-limited (Gaussian) magnetic contribution to the scattering and the blue area represents the dynamic, quasi-elastic or frozen (Lorentzian) contributions to the scattering. The red area shows the structural scattering obtained via a fit of the 3T data set. The black lines shows the resulting fit which are obtained by the addition of the three curves (blue,yellow and red). }
\label{FitBragg}
\end{figure*}

To extract the long-range order as well as the dynamic, quasi-elastic and frozen contributions to the (220) magnetic Bragg peak, we performed an integration of the neutron scattering data set of $-0.2$ to 0.2 in the [00L] direction and $-0.1$ to 0.1 meV in energy transfer. The results of such integration for the three single crystals are shown in Fig.~\ref{FitBragg} for $T=0.1$~K and zero applied field. Each Bragg peak has been fit to a Gaussian function to account for the long-range order (blue shaded region in Fig.~\ref{FitBragg}) and a Lorentzian function for the dynamic, quasi-elastic and frozen contributions to the scattering (yellow shaded region in Fig.~\ref{FitBragg}). The width of the Gaussian function was fixed by the width of a strictly nuclear reflection in the 3T data set. An additional Gaussian contribution to the scattering was accounted for by the 3T data set itself, representing the nuclear contribution (red shaded region in Fig.~\ref{FitBragg}) and kept fix for the low field data sets.  Finally, the width of the Lorentzian function was allowed to refine freely. The results for all samples show that the relative contribution of the Lorentzian scattering over the Gaussian (long-range order) contributions is strongly enhanced as a function of doping. As discussed in the main manuscript, this indicates that upon dilution, the spins do not order in the discrete states belonging to $\psi_2$ but instead freeze into a mosaic of $\psi_2$ and $\psi_3$ domains, whose characteristic size is approximated by the inverse of the $Q$-width of the Lorentzian component to the scattering, which gives 35 $\pm$ 7 \AA.

\section{Temperature dependence of the (220) Bragg peak for 20\%-Y E\lowercase{r}$_{2}$T\lowercase{i}$_2$O$_7$.}

\begin{figure}[tbp]
\linespread{1}
\par
\includegraphics[width=2.8in]{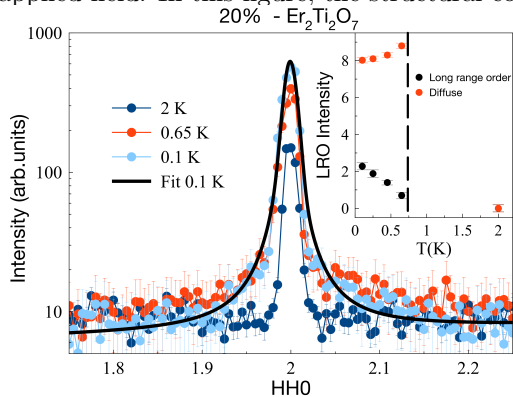}
\par
\caption{Elastic cut showing the (220) magnetic Bragg peak for varying temperature in Er$_{2-x}$Y$_x$Ti$_2$O$_7$ with $x$=0.4 (20$\%$-Y) in zero applied field. The black line represents the resulting fit for the $T=0.1$~K data set. The inset panel shows the temperature dependence of the Gaussian, long range ordered, and Lorentzian contributions to the (220) scattering, where the dashed line marks the N\'eel transition temperature, $T_N=0.74(8)$~K.}
\label{fig5}
\end{figure}

The temperature dependence of the (220) magnetic Bragg peak in Er$_{2-x}$Y$_x$Ti$_2$O$_7$ with $x=0.4$ (20$\%$-Y) has also been explored to investigate a possible phase transition between $\psi_2$ and $\psi_3$ states at temperatures between the base temperature of our neutron experiment ($T=0.1$~K) and the N\'eel temperature ($T=0.74(8)$~K) of this sample. Fig.~\ref{fig5} shows elastic cuts around the (220) Bragg position obtained by an integration of $-0.2$ to 0.2 in the [00L] direction and $-0.1$ to 0.1 meV in energy transfer. The same procedure followed in the previous section was again employed to fit the (220) magnetic Bragg peak at every temperature collected in our neutron experiment in zero applied field. In this figure, the structural contribution of the long-range order scattering at (220) has been removed using intensity obtained from the paramagnetic, $T=2$~K data set. The temperature dependence of the Lorentzian and the long-range order scattering are shown in the inset of Fig.~\ref{fig5}. The long-range order contribution shows typical order parameter behavior, correlated with $T_N$ and no anomaly is observed below that temperature. In contrast, the Lorentzian scattering has little temperature dependence but is enhanced near $T_N$, consistent with critical scattering near a continuous phase transition.

\bibliography{EYTO}